\documentclass[aps,prl,twocolumn,showpacs]{revtex4}
\usepackage{bm}
\usepackage{amsmath}
\usepackage{amssymb}
\usepackage{graphicx}

\newcommand{\nb}{{\bf n}}
\newcommand{\lan}{\langle}
\newcommand{\ran}{\rangle}
\newcommand{\Sig}{\Sigma}

\newcommand{\Ga}{\Gamma}
\newcommand{\eps}{\varepsilon}

\newcommand{\gapara}{\gamma_\text{para}}
\newcommand{\gaorb}{\gamma_\text{orb}}

\begin{document}
\noindent {\bf Comment on ``Magnetic-Field Enhancement of Superconductivity in Ultranarrow Wires''}\\

In a recent Letter \cite{themPRL}, the authors observed enhancement 
of the critical supercurrent $I_c(H,T)$ of superconducting nanowires
in external magnetic field $H$. 
They explain this behavior using
the mechanism of enhancement of superconductivity(SC) 
due to presence of {\em polarized magnetic moments} in the samples, 
the theory of which was developed by us in Ref.~\cite{us}. 
The authors of Ref.~\cite{themPRL} adopted
our theory to derive Gor'kov equations
in Ref.~\cite{themEPL}, and, by solving them numerically, obtained $I_c(H,T)$. 
Indeed, good fitting of experimental to numerical curves was obtained
in Ref.~\cite{themPRL}.  

In Ref.~\cite{us} we showed that 
two quite stringent conditions 
on the smallness of {\em orbital}(OE) and {\em paramagnetic}(PE) {\em effects}
must be met in order to observe the enhancement of SC. 
These conditions, 
despite of  being a crucial part of theoretical analysis of the effect, 
were not derived in Refs.~\cite{themEPL,themPRL}
and, consequently, were not discussed in Ref.~\cite{themPRL}
in connection with the experiment. 
We reproduce these conditions 
for the case of wires below
and show that they explain 
the key features of the experiment~\cite{themPRL} analytically.

Essential physics of the effect \cite{us} is contained in the self-energy
(see Eq.~(8) for the Cooperon in Ref.~\cite{us})
\[
\Sig(\eps,H,T)=\Ga(\eps)+\gaorb(H)+\gapara(H,T)
\]
of the anomalous Green function, 
given by the sum of  
(i) the {\em total} (``spin-flip''+``non-spin-flip'') 
rate $\Ga(\eps)=\Ga(\eps,H,T)$ of exchange scattering(ES) on magnetic impurities,
(ii) orbital $\gaorb(H) \propto \theta_\nb D(\frac{e}{c}Hd)^2$
and (iii) paramagnetic $\gapara(H,T) \propto \tau_\text{so} h'^2$ 
depairing rates. 
Here $D$ is the diffusion constant, $d$ is the wire diameter, 
$\tau_\text{so}$ is the spin-orbit scattering time,
$h'=\mu_B H - n_S J \lan S_z \ran$ is the exchange field.
The factor $\theta_\nb \sim 1$ depends on the direction $\nb$ of 
magnetic field $H$ and reaches its  minimum $\theta_\parallel$ for parallel 
and maximum $\theta_\perp$ for perpendicular orientations of $H$ relative to the wire,
$\theta_\parallel \leq \theta_\nb \leq \theta_\perp$.
The enhancement of SC is due to the {\em decrease} of $\Ga(\eps)$ 
from $\Ga(\eps)=\frac{1}{\tau_S}$ for unpolarized impurities at $H=0$ to
$\Ga(\eps)=\frac{1}{\tau_S} \frac{S}{S+1}$ for strongly polarized impurities at $\mu_B H \gg T$.
Note that the effect is greatest for $S=1/2$ and vanishes for $S\gg 1$~\cite{us}.

The current $I_c(H,T)$ at a given temperature $T$ or 
the transition temperature $T_c(H)$ (such that $T_c(0)=T$)
of a wire are nonmonotonic (first growing, then decreasing) in $H$
{\em if and only if}
$\Sig(\eps,H,T)$
is a decreasing function of $H$ at a given $T$ 
for  small fields $\mu_B H \ll T$.
Using the expressions for $\Ga(\eps)$, $\gaorb(H)$, $\gapara(H,T)$, \cite{us},
we find that the effect is well-pronounced 
if the conditions
\begin{eqnarray}
	1/(\tau_S T^2) \gg \theta_\nb D m^2 d^2 \label{OE} \\
	1/\tau_S \gg \tau_\text{so} (n_S J)^2 \label{PE}
\end{eqnarray}
for the smallness of $\gaorb$ (OE) and $\gapara$ (PE)
compared to the decrease $\frac{1}{\tau_S}-\Ga(\eps)$ in ES rate, 
respectively, are met (cf. Eqs.~(2),(3) of Ref.~\cite{us}).
In Eqs.~(\ref{OE}),(\ref{PE}) $m$ is electron mass, $S\sim 1$ is assumed,
$\hbar=1$.

Note that Eq.~(\ref{OE}) is $T$-dependent,
whereas Eq.~(\ref{PE}) is not. 
Therefore 
(i) the relevance of OE (compared to ES) varies with $T$:
the lower $T$ is, the better Eq.~(\ref{OE}) is satisfied, and hence,
the less significant OE is;
at higher $T$ OE eventually dominates over ES;
(ii) on the contrary, 
the strength of PE (relative to ES) is the same for both $T\ll T_c(0)$
and $T\sim T_c(0)$: since the ``polarization factor'' $\mu_B H/T$
enters both $\frac{1}{\tau_S}-\Ga(\eps)$ and $\gapara(H,T)$,
the temperature $T$ eventually drops out of Eq.~(\ref{PE}).

We come to an important conclusion.  
Qualitative difference in the behavior 
of $I_c(H,T)$ at $T=0.3\text{K} \ll T_c(0)$ (nonmonotonic in $H$) 
and at $T \approx 2-4\text{K} \sim T_c(0)$ (monotonic in $H$)
is due to the competition of the decreasing ES rate
with {\em orbital effect alone}, but not with paramagnetic effect.
Therefore, diminishing
of the anomaly in $I_c(H,T)$ with
growing $T$ in Fig.~3(a) of Ref.~\cite{themPRL} 
should be attributed to OE only.
Indeed, inserting numerical values provided in Ref.~\cite{themPRL}, 
we check that Eq.~(\ref{PE}) is well satisfied for all samples 
($\text{LHS/RHS} \sim 10$). 
At the same time we check that at  $T=0.3$K Eq.~(\ref{OE}) is very well satisfied 
($\text{LHS/RHS} \sim 100$)
for all MoGe wires, 
whereas at $T \approx 2-4$K the condition (\ref{OE}) 
is violated ($\text{LHS/RHS} \sim 1$).
Another interesting feature, that $T_c(H)$ of the wire monotonically decreases with $H$,
although $I_c(H,T)$ at $T=0.3$K is nonmonotonic, 
is also explained by the fact that Eq.~(\ref{OE}) is violated for $T=T_c(0) \approx 2-4$K.

The relevance of the value of wire diameter $d$ (Fig.~2(b))
and orientation $\nb$ of magnetic field (Fig.~2(c))
is  clearly seen from Eq.~(\ref{OE}). 
Magnetic field $H^*$, at which OE eventually overtakes decreasing ES rate,
can be estimated as $1/\tau_S \sim \theta_\nb D(\frac{e}{c}H^*d)^2$.
Since $\theta_\parallel < \theta_\perp$ one gets $H^*_\parallel > H^*_\perp$
(Fig.~2(c)).
Behavior of $I_c(H,T)$ at small fields $\mu_B H \ll T$ is insensitive to $\nb$
(Fig.~2c), only because Eq.~(\ref{OE}) is {\em very well} satisfied
at $T=0.3K$; for greater $T$ this low-field behavior 
would be more sensitive to $\nb$.

To conclude, simple estimates (\ref{OE}),(\ref{PE}) following from our theory \cite{us} 
are shown to be the crucial elements of the analysis of experiment~\cite{themPRL}.
\\
\\
\noindent Maxim Yu. Kharitonov$^{1}$ and Mikhail V. Feigel'man$^{2}$\\
{\small $^{1}$TP III,
Ruhr Universit\"{a}t Bochum, Germany,\\
$^{2}$Landau Institute for Theoretical Physics,
Moscow, Russia.

}

\end{document}